\documentstyle[11pt]{article}
\def\cE{\cal E}

\def\cL{\cal L}

\newfont{\goth}{eufm10 scaled \magstep1}

\def\a{\alpha}
\def\b{\beta}
\def\c{\gamma}\def\C{\Gamma}
\def\d{\delta}
\def\e{\epsilon}

\def\h{\eta}
\def\k{\kappa}

\def\m{\mu}

\def\P{\Pi}

\def\th{\theta}

\def\beq{\begin{equation}}\def\eeq{\end{equation}}
\def\beqa{\begin{eqnarray}}\def\eeqa{\end{eqnarray}}
\def\barr{\begin{array}}\def\earr{\end{array}}

\def\o{\omega}\def\O{\Omega}
\def\del{\partial}
\def\ua{\underline{\alpha}}
\def\ub{\underline{\phantom{\alpha}}\!\!\!\beta}
\def\uc{\underline{\phantom{\alpha}}\!\!\!\gamma}

\def\ud{\underline\delta}

\def\una{\underline a}\def\unA{\underline A}
\def\unb{\underline b}\def\unB{\underline B}
\def\unc{\underline c}\def\unC{\underline C}
\def\und{\underline d}
\def\une{\underline e}

\def\unm{\underline m}\def\unM{\underline M}
\def\unn{\underline n}

\def\nab{\nabla}

\let\la=\label

\let\bm=\bibitem

\def\umu{{\underline \mu}}
\def\nn{\nonumber}
\def\bd{\begin{document}}
\def\ed{\end{document}}
\def\ba{\begin{array}}
\def\ea{\end{array}}
\def\bea{\begin{eqnarray}}
\def\eea{\end{eqnarray}}
\def\ft#1#2{{\textstyle{{\scriptstyle #1}\over {\scriptstyle #2}}}}
\def\fft#1#2{{#1 \over #2}}
\newcommand{\be}{\begin{equation}}
\newcommand{\ee}{\end{equation}}
\newcommand{\eq}[1]{(\ref{#1})}
\def\eqs#1#2{(\ref{#1}-\ref{#2})}
\def\det{{\rm det\,}}
\def\tr{{\rm tr}}
\newcommand{\ho}[1]{$\, ^{#1}$}
\newcommand{\hoch}[1]{$\, ^{#1}$}
\newcommand{\tamphys}{\it\small Center for Theoretical Physics, 
Texas A\&M University, College Station, TX 77843, USA}
\newcommand{\newton}{\it\small Isaac Newton Institute for Mathematical Sciences,
Cambridge, UK}
\newcommand{\kings}{\it\small Department of Mathematics, King's College, London, UK}

\newcommand{\auth}{\large P.S. Howe\hoch{1}, E. Sezgin \hoch{2\dagger} 
and P.C. West \hoch{3\ddagger}}

\begin{document}

\hfill{KCL-TH-97-05}

\hfill{CTP TAMU-8/97}

\hfill{NI-97002}

\hfill{hep-th/9702008}

\vspace{20pt}

\begin{center}
{\Large Covariant Field Equations of the M Theory Five-Brane}
\vspace{30pt}

\auth

\vspace{15pt}

\begin{itemize}
\item [$^1$] \kings
\item [$^2$] \tamphys
\item [$^3$] \newton
\end{itemize}

\vspace{60pt}

{\bf Abstract}

\end{center}

The component form of the equations of motion for the 5-brane in
eleven-dimensions is derived from the superspace equations. These equations
are fully covariant in six-dimensions. It is shown
that double-dimensional reduction of the bosonic equations gives the
equations of motion for a 4-brane in ten dimensions governed by the
Born-Infeld action.

{\vfill\leftline{}\vfill
\vskip	10pt
\footnoterule
{\footnotesize
\hoch{\dagger} Research supported in part by NSF Grant	PHY-9411543 \vskip -12pt} 
\vskip	10pt 
{\footnotesize
\hoch{\ddagger} Permanent Address: \kings \vskip -12pt}}

\pagebreak
\setcounter{page}{1}

\section{Introduction}

It is now widely believed that there is a single underlying theory which
incorporates all superstring theories and which also has, as a
component, a new theory in eleven-dimensions which has been christened
``$M$-theory''. Opinion is divided as to whether $M$-theory is itself
the fundamental theory or whether it is one corner of a large moduli
space which has the five consistent ten-dimensional superstring theories
as other corners. Whichever viewpoint turns out to be correct it seems
certain that $M$-theory will play a crucial r\^{o}le in future
developments. Not much is known about this theory at present, apart from
the fact that it has eleven-dimensional supergravity as a low energy
limit and that it has two basic BPS $p$-branes, the 2-brane and the
5-brane, which preserve half-supersymmetry. The former can be viewed as
a fundamental (singular) solution to the supergravity equations whereas
the latter is solitonic. It is therefore important to develop a better
understanding of these branes and in particular the 5-brane, since the
Green-Schwarz action for the 2-brane has been known for some time.

In a recent paper \cite{hs1} it was shown that all branes preserving
half-supersymmetry can be understood as embeddings of one superspace,
the worldsurface, into another, the target superspace, which has
spacetime as its body, and that the basic embedding condition which
needs to be imposed is universal and geometrically natural. The results
of \cite{hs1} were given mainly at the linearised level; in a sequel
\cite{hs2} the eleven-dimensional 5-brane was studied in more detail and
the full non-linear equations of motion were derived. However, these
were expressed in superspace notation. It is the purpose of this paper
to interpret these equations in a more familiar form, in other words to
derive their component equivalents. In the context of superembeddings
the component formalism means the Green-Schwarz formalism since the
leading term in the worldsurface $\th$-expansion of the embedding
describes a map from a bosonic worldsurface to a target superspace.

Partial results for the bosonic sector of the eleven-dimensional
fivebrane have been obtained in \cite{f1,f2,f3,f4}. More recently, a
non-covariant bosonic action has been proposed \cite{f5,f6}. In this
approach, only the five-dimensional covariance is manifest. In \cite{f7},
a complete bosonic action has been constructed. The action contains an
auxiliary scalar field, which can be eliminated at the expense of
sacrificing the six dimensional covariance, after which it reduces to 
the action of \cite{f5,f6}.

In this paper we will show that the covariant superfield equations of
motion of the eleven-dimensional superfivebrane presented in \cite{hs1,hs2}
can be written in $\kappa$-invariant form, and that they do have the anticipated
Born-Infeld structure. The $\kappa$-symmetry emerges from the worldsurface
diffeomorphism invariance of the superspace equations, the parameter of
this symmetry being essentially the leading component in the worldsurface
$\th$-expansion of an odd diffeomorphism. We
find neither the need to introduce a scalar auxiliary field, nor the
necessity to have
only five-dimensional covariance. As long as one does not insist on
having an action, it is possible to write down six dimensional covariant
equations, as one normally expects in the case of chiral $p$-forms.

In order to show that our equations have the expected Born-Infeld form
we perform a double-dimensional reduction and compare them with the
equations of motion for a 4-brane in ten dimensions. In section 4, we
do this comparison in the bosonic sector, and flat target space, and
show that  the Born-Infeld form of the 4-brane equations of
motion does indeed emerge. The work of refs \cite{hs1,hs2} is briefly
reviewed in the next section,
and in section 3 the equations of motion are described in Green-Schwarz
language.

\section{Equations of Motion in Superspace}

The 5-brane is described by an embedding of the worldsurface $M$, which
has (even$|$odd) dimension $(6|16)$ into the target space, $\unM$, which
has dimension $(11|32)$. In local coordinates
$z^{\unM}=(x^{\unm},\th^{\umu})$ for $\unM$ and $z^M$ for $M$ the
embedded submanifold is given as $z^{\unM}(z)$\footnote{We shall also
denote the coordinates of $M(\unM)$ by $z=(x,\th)({\underline
z}=(\underline{x},\underline{\th}))$ if it is not necessary to use
indices}. We define the embedding matrix $E_A{}^{\unA}$ to be the
derivative of the embedding referred to preferred bases on both
manifolds:
\beq
E_A{}^{\unA}=E_A{}^M\del_{M}z^{\unM}E_{\unM}{}^{\unA},
\eeq
where $E_M{}^A\ (E_A{}^M)$ is the supervielbein (inverse supervielbein)
which relates the preferred frame basis to the coordinate basis, and the
target space supervielbein has underlined indices. The notation is as
follows: indices from the beginning (middle) of the alphabet refer to
frame (coordinate) indices, latin (greek) indices refer to even (odd)
components and capital indices to both, non-underlined (underlined)
indices refer to $M\ (\unM)$ and primed indices refer to normal
directions. We shall also employ a two-step notation for spinor indices;
that is, for general formulae a spinor index $\a$ (or $\a'$) will run
from 1 to 16, but to interpret these formulae, we shall replace a
subscript $\a$ by a subscript pair $\a i$ and a subscript $\a'$ by a
pair ${}_i^{\a}$, where $\a=1,\dots 4$ and $i=1,\dots 4$ reflecting the
$Spin(1,5)\times USp(4)$ group structure of the $N=2,d=6$ worldsurface
superspace. (A lower (upper) $\a$ index denotes a left-handed
(right-handed) $d=6$ Weyl spinor and the $d=6$ spinors that occur in the
theory are all symplectic Majorana-Weyl.)

The torsion 2-form $T^{\unA}$ on $\unM$ is given as usual by
\beq
T^{\unA}=d E^{\unA} +E^{\unB} \O_{\unB}{}^{\unA}
\eeq
where $\O$ is the connection 1-form. The pull-back of this equation onto
the worldsurface reads, in index notation,
\beq
\nabla_A E_B{}^{\unC}-(-1)^{AB}\nabla_{B} E_{A}{}^{\unC} +T_{AB}{}^C
E_C{}^{\unC}
=(-1)^{A(B+\unB)}E_B{}^{\unB} E_A{}^{\unA} T_{\unA\unB}{}^{\unC}\ , \label{de}
\eeq
where the derivative $\nabla_A$ is covariant with respect to both
spaces, i.e. with respect to both underlined and non-underlined indices,
the connection on $M$ being, at this stage at least, independent of the
target space connection.

The basic embedding condition is
\beq
E_{\a}{}^{\una}=0\ , \label{mc1}
\eeq
from which it follows that (using \eq{de})
\beq
E_\a{}^{\ua}E_\b{}^{\ub}T_{\ua\ub}{}^{\unc}=T_{\a\b}{}^cE_c{}^{\unc}\
.\label{mc2}
\eeq
If the target space geometry is assumed to be that of (on-shell)
eleven-dimensional supergravity equation \eq{mc1} actually determines
completely the induced geometry of the worldsurface and the dynamics of
the 5-brane. In fact, as will be discussed elsewhere, it is not
necessary to be so specific about the target space geometry, but it will
be convenient to adopt the on-shell geometry in the present paper. The
structure group of the target superspace is $Spin(1,10)$ and the
non-vanishing parts of the target space torsion are \cite{cf,bh}
\beq
T_{\ua\ub}{}^{\unc}=-i(\C^{\unc})_{\ua\ub}\ ,
\eeq
\beq
T_{\una\ub}{}^{\uc}=-
{1\over36}(\C^{\unb\unc\und})_{\ub}{}^{\uc}H_{\una\unb\unc\und}
-{1\over288}(\C_{\una\unb\unc\und\une})_{\ub}{}^{\uc}H^{\unb\unc\und\une}\ ,
\la{tt}
\eeq
where $H_{\una\unb\unc\und}$ is totally antisymmetric, and the dimension 3/2
component $T_{\una\unb}{}^{\uc}$. $H_{\una\unb\unc\und}$ is the dimension one
component of the closed superspace 4-form $H_4$ whose only other non-vanishing
component is
\beq
H_{\una\unb\uc\ud}=-i(\C_{\una\unb})_{\uc\ud}\ .
\eeq
With this target space geometry equation \eq{mc2} becomes
\beq
E_\a{}^{\ua}E_\b{}^{\ub}(\C^{\unc})_{\ua\ub}=iT_{\a\b}{}^cE_c{}^{\unc}
\label{me}\ .
\eeq

The solution to this equation is given by
\beq
E_{\a}{}^{\ua}=u_{\a}{}^{\ua}+h_{\a}{}^{\b'} u_{\b'}{}^{\ua}\ , \la{s}
\eeq
and
\beq
E_{a}{}^{\una}=m_a{}^b u_{b}{}^{\una} ,  \label{m}
\eeq
together with
\beq
T_{\a\b}{}^c=-i(\C^c)_{\a\b}\rightarrow -i\h_{ij}(\c^c)_{\a\b}\ .
\eeq
with $\h_{ij}=-\h_{ji}$ being the $USp(4)$ invariant tensor and the pair
$(u_\a{}^{\ua},u_{\a'}{}^{\ua})$ together making up an element of the
group $Spin(1,10)$. Similarly, there is a $u_{a'}{}^{\una}$ such that
the pair $(u_a{}^{\una},u_{a'}{}^{\una})$ is the element of $SO(1,10)$
corresponding to this spin group element. (The inverses of these group
elements will be denoted $(u_{\ua}{}^{\a},u_{\ua}{}^{\a'})$ and
$(u_{\una}{}^a,u_{\una}{}^{a'})$.) The tensor $h_\a{}^{\b'}$ is
given by \footnote{We have rescaled the $H_{abc}$ and $h_{abc}$ of
refs \cite{hs1,hs2} by a factor of 6.}
\beq
h_{\a}{}^{\b'}\rightarrow h_{\a i \b}^{\phantom{\a i} j}=
{1\over 6} \d_i{}^j(\c^{abc})_{\a\b} h_{abc}\ ,
\eeq
where $h_{abc}$ is self-dual, and
\beq
m_a{}^b=\d_a{}^b-2h_{acd} h^{bcd}\ . \la{mmm}
\eeq
This solution is determined up to local gauge transformations belonging
to the group $Spin(1,5)\times USp(4)$, the structure group of the
worldsurface. One also has the freedom to make worldsurface super-Weyl
transformations but one can consistently set the conformal factor to be
one and we shall do this throughout the paper.

It is useful to introduce a normal basis
$E_{A'}=E_{A'}{}^{\unA} E_{\unA}$ of vectors at each point on the
worldsurface. The inverse of the pair $(E_A{}^{\unA},E_{A'}{}^{\unA})$
is denoted by $(E_{\unA}{}^A,E_{\unA}{}^{A'})$.
The odd-odd and
even-even components of the normal matrix $E_{A'}{}^{\unA}$ can be
chosen to be
\beq
E_{\a'}{}^{\ua}=u_{\a'}{}^{\ua}\ , \la{c1}
\eeq
and
\beq
E_{a'}{}^{\una}=u_{a'}{}^{\una}\ . \la{c2}
\eeq
Together with \eq{s} and \eq{m}, it follows that the inverses
in the odd-odd and even-even sectors are
\beq
E_{\ua}{}^{\a}=u_{\ua}{}^{\a}\ , \qquad
E_{\ua}{}^{\a'}=u_{\ua}{}^{\a'}-u_{\ua}{}^{\b} h_{\b}{}^{\a'}\ , \la{ss1}
\eeq
and
\beq
E_{\una}{}^a=u_{\una}{}^b(m^{-1})_b{}^a\ ,\qquad
E_{\una}{}^{a'}=u_{\una}{}^{a'}\ . \la{ss2}
\eeq
Later, we will also need the relations \cite{hs2}
\beqa
u_{\a}{}^{\ua} u_{\b}{}^{\ub} (\C^{\una})_{\ua\ub} &=&
(\C^{a})_{\a\b}u_{a}{}^{\una}\ , \la{u1}\\
u_{\a'}{}^{\ua} u_{\b'}{}^{\ub} (\C^{\una})_{\ua\ub} &=&
(\C^{a})_{\a'\b'}u_{a}{}^{\una}\ , \la{u2}\\
u_{\a}{}^{\ua} u_{\b'}{}^{\ub} (\C^{\una})_{\ua\ub} &=&
(\C^{a'})_{\a\b'}u_{a'}{}^{\una}\ , \la{u3}
\eeqa
which follow from the fact that the $u$'s form a $32\times 32$
matrix that is an element of $Spin(1,10)$.

The field $h_{abc}$ is a self-dual antisymmetric tensor, but it is not
immediately obvious how it is related to a 2-form potential. In fact, it
was shown in \cite{hs2} that there is a superspace 3-form $H_3$ which
satisfies
\beq
dH_3=-{1\over 4}H_4 ,
\eeq
where $H_4$ is the pull-back of the target space 4-form, and whose only
non-vanishing component is $H_{abc}$ where
\beq
H_{abc}=m_a{}^d m_b{}^e h_{cde}.  \label{h}
\eeq

The equations of motion of the 5-brane can be obtained by systematic
analysis of the torsion equation \eq{de}, subject to the condition
\eq{mc1} \cite{hs2}. The bosonic equations are the scalar equation
\beq
\h^{ab} K_{a b}{}^{c'}={1\over8}(\c^{c'})^{jk}(\c^a)^{\b\c}Z_{a,\b j,\c k}\ ,
\label{dil}
\eeq
and the antisymmetric tensor equation
\beq
\hat\nabla^c h_{abc}=-{\h^{jk}\over 16}\big( (\c_{[a})^{\b\c}Z_{b],\b j,\c k}
+{1\over2}(\c_{ab}{}^c)^{\b\c}Z_{c,\b j,\c k}\big)\ , \label{tensor}
\eeq
where
\beq
Z_{a\b}{}^{\c'}=E_{\b}{}^{\ub} E_a{}^{\una} T_{\una\ub}{}^{\uc}E_{\uc}{}^{\c'}
-E_a{}^{\uc}\nabla_{\b} E_{\uc}{}^{\c'}\ , \label{z}
\eeq
and
\beq
\hat\nabla_a h_{bcd}=\nabla_a h_{bcd}-3X_{a,[b}{}^e h_{cd]e}\ ,
\eeq
with
\beq
X_{a,b}{}^c=(\nabla_a u_b{}^{\unc}) u_{\unc}{}^{c}\ . \la{xx}
\eeq
In the scalar equation we have introduced a part of the second
fundamental form of the surface which is defined to be
\beq
K_{AB}{}^{C'}=(\nabla_A E_B{}^{\unC})E_{\unC}{}^{C'}\ .\label{sff}
\eeq

Finally, the spin one-half equation is simply
\beq
(\c^a)^{\a\b}\chi_{a\b}^{\phantom{a}j}=0\ ,\label{Dirac}
\eeq
where
\beq
\chi_a{}^{\a'}=E_a{}^{\ua} E_{\ua}{}^{\a'}\ . \la{chi}
\eeq

We end this section by rewriting the equations of motion \eq{dil},
\eq{tensor} and \eq{Dirac} in an alternative form that will be useful
for the purposes of the next section:
\beqa
&& E_a{}^{\ua} E_{\ua}{}^{\b'}(\C^a)_{\b'}{}^{\a}=0\ , \la{Dirac2}\\
&&\nn\\
&&\eta^{ab}\nabla_a E_b{}^{\una} E_{\una}{}^{b'} =
-\ft18 (\C^{b'a})_{\c'}{}^{\b}\, Z_{a\b}{}^{\c'} \ ,\la{dil2}\\
&&\nn\\
&&\hat\nabla^c h_{abc}=-\ft1{32}(\C^c\C_{ab})_{\c'}{}^{\b} Z_{c\b}{}^{\c'}\
. \label{tensor2}
\eeqa
It will also prove to be useful to rewrite \eq{z} as
\beq
Z_{a\b}{}^{\c'}=E_\b{}^{\ub}~\left(T_{a\ub}{}^{\uc}-
K_{a\ub}{}^{\uc}\right)~E_{\uc}{}^{\c'}\ , \la{z2}
\eeq
with the matrices $T_a$ and $K_a$ defined as
\beqa
T_{a\ub}{}^{\uc} &=& E_a{}^{\una} T_{\una\ub}{}^{\uc}\ , \la{t2}\\
K_{a\ub}{}^{\uc} &=& E_a{}^{\ud} E_{\ub}{}^\c
(\nabla_\c E_{\ud}{}^{\d'}) E_{\d'}{}^{\uc}\ . \la{k}
\eeqa

\section{Equations of Motion in Green-Schwarz Form}

\subsection{Preliminaries}

In this section we derive the component equations of motion
following from the superspace equations given in the last section. The idea
is to expand the superspace
equations as power series in $\th^{\m}$ and to evaluate them at $\th=0$.
We may choose a gauge in which the worldsurface supervielbein takes the
form
\beq
\barr{lcllcl}
E_m{}^a(x,\th) &=&E_m{}^a(x) + O(\th)\ \
& E_m{}^{\a}(x,\th)&=&E_m{}^{\a}(x)+O(\th)\\
E_{\m}{}^a(x,\th)&=&0+O(\th)\ \
& E_{\m}{}^{\a}(x,\th)&=&\d_{\m}{}^{\a}+O(\th),
\earr
\eeq
and the inverse takes the form
\beq
\barr{lcllcl}
E_a{}^m(x,\th) &=&E_a{}^m(x) + O(\th)\ \
& E_a{}^{\m}(x,\th)&=&E_a{}^{\m}(x)+O(\th)\\
E_{\a}{}^m(x,\th)&=&0+O(\th)\ \
& E_{\a}{}^{\m}(x,\th)&=&\d_{\a}{}^{\m}+O(\th),
\earr
\eeq
where $E_a{}^m(x)$ is the inverse of $E_m{}^a(x)$. The component field
$E_m{}^{\a}(x)$ is the worldsurface gravitino, which is determined by
the embedding, but which only contributes terms to the equations of
motion which we shall not need for the purpose of this section. The
field $E_a{}^{\m}(x)$ is linearly related to the gravitino. From the
embedding condition \eq{mc1} we learn that
\beq
\del_{\m}z^{\unM}E_{\unM}{}^{\una}=0\qquad {\rm at}\  \th=0,
\eeq
so that
\beqa
E_a{}^{\una} &=& E_a{}^m {\cE}_m{}^{\una}\  \qquad {\rm at}\  \th=0\ , \la{ep}\\
E_a{}^{\ua} &=& E_a{}^m {\cE}_m{}^{\ua}\  \qquad {\rm at}\  \th=0\ ,
\eeqa
where we have used the definitions
\beqa
{\cE}_m{}^{\una}(x) &=& \del_m z^{\unM} E_{\unM}{}^{\una}\qquad {\rm at}\
\th=0\ ,\\
{\cE}_m{}^{\ua}(x) &=& \del_m z^{\unM} E_{\unM}{}^{\ua}\qquad {\rm at}\ \th=0\ .
\eeqa
These are the embedding matrices in the Green-Schwarz formalism, often
denoted by $\P$. From \eq{m} we have
\beq
E_a{}^{\una} E_b{}^{\unb}\h_{\una\unb}=m_a{}^c m_b{}^d\h_{cd},
\eeq
this equation being true for all $\th$ and in particular for $\th=0$.
Therefore, if we put
\beq
e_a{}^m=((m^{-1})_a{}^b E_b{}^m)(x) , \label{e}
\eeq
we find that $e_m{}^a$ is the sechsbein associated with the standard
$GS$ induced metric
\beq
g_{mn}(x)={\cE}_m{}^{\una}{\cE}_n{}^{\unb}\h_{\una\unb}.
\eeq

There is another metric, which will make its appearance later, which
we define as
\beqa
G^{mn}&=& E_a{}^m(x) E_b{}^n(x)\, \eta^{ab}  \\
      &=& ((m^2)^{ab}e_a{}^m e_b{}^n)(x)\ . \la{gg}
\eeqa
We also note the relation
\beq
u_a{}^{\una} = e_a{}^m {\cE}_m{}^{\una}\ , \la{ue}
\eeq
which follows from \eq{m}, \eq{ep} and \eq{e}.

For the worldsurface 3-form $H_3$ we have
\beq
H_{MNP}=E_P{}^C E_B{}^N E_A{}^M H_{ABC}(-1)^{((B+N)M+(P+C)(M+N))}
\eeq
Evaluating this at $\th=0$ one finds
\beq
H_{mnp}(x)=(E_m{}^a E_n{}^b E_p{}^c H_{abc})(x)
\eeq
so that, using \eq{h} and \eq{e}, one finds
\beq
h_{abc}(x)=m_a{}^d e_d{}^m e_b{}^n e_c{}^p H_{mnp}(x)\ . \la{hh}
\eeq

We are now in a position to write down the equations of motion in terms
of ${\cE}_m{}^{\ua}$, ${\cE}_m{}^{\una}$ and $H_{mnp}(x)$. The basic
worldsurface fields are $x^{\unm}, \theta^{\underline\m}$ and $B_{mn}(x)$, where
$B_{mn}$ is the 2-form potential associated with $H_{mnp}$ as
$H_3=dB_2-\ft14 C_3$ and $C_3$ is the pull-back of the target space
3-form. We begin with the Dirac equation \eq{Dirac}.

\subsection{The Dirac Equation}

In order to extract the Dirac equation in $\kappa$-invariant component
form, it is convenient to define the projection operators
\beqa
E_{\ua}{}^{\a} E_{\a}{}^{\uc} &=& \ft12 (1+\C)_{\ua}{}^{\uc}\ ,
\la{p1}\\
E_{\ua}{}^{\a'} E_{\a'}{}^{\uc} &=& \ft12 (1-\C)_{\ua}{}^{\uc}\ .  \la{p2}
\eeqa
The $\C$-matrix, which clearly satisfies $\C^2=1$, can be calculated from these
definitions as follows. We expand
\beq
E_{\ua}{}^{\a} E_{\a}{}^{\uc}=\sum_{n=0}^5 C^{\una_1\cdots\una_n}
\left(\C_{\una_1\cdots\una_n}\right)_{\ua}{}^{\uc}\ , \la{exp}
\eeq
where $C$'s are the expansion coefficients that are to be determined.
Tracing this equation with suitable $\C$-matrices, and using the
relations \eqs{u1}{u3}, we find that the only non-vanishing coefficients are
\beqa
C &=& {1\over 2}\ ,\\
C^{\una\unb\unc} &=& {1\over 6} h^{abc}
u_a{}^{\una}u_b{}^{\unb}u_c{}^{\unc}\ ,\\
C^{\una_1\cdots\una_6} &=& -{1\over 6!2}\e^{a_1\cdots
a_6}u_{a_1}{}^{\una_1}\cdots
u_{a_6}{}^{\una_6}\ . \la{co}
\eeqa
Substituting these back, and comparing with \eq{p1}, we find
\beq
\C={1\over 6!\sqrt {-g}} \e^{m_1\cdots m_6} \left(- \C_{m_1\cdots m_6}
+40 \C_{m_1\cdots m_3} h_{m_4\cdots m_6}\right)\ ,\la{cc}
\eeq
where we have used \eq{ue} and the definitions
\beqa
\C_m &=& {\cE}_m{}^{\una}\C_{\una} \ . \la{pb}\\
h_{mnp} &=& e_m{}^a e_n{}^b e_p{}^c h_{abc}\ .
\eeqa

The matrix $\C$ can also be written as
\beq
\C=(-1+\ft13\C^{mnp}h_{mnp})\C_{(0)}\ , \la{af}
\eeq
where
\beq
\C_{(0)}={1\over 6!\sqrt {-g}} \e^{m_1\cdots m_6} \C_{m_1\cdots m_6}\ . \la{c0}
\eeq

It is now a straightforward matter to derive the component for of the
Dirac equation \eq{Dirac2}. We use \eq{u1} to replace the worldsurface
$\C$-matrix by the target space $\C$-matrix multiplied by factors of
$u$, and recall \eq{c1}, \eq{ss1}, \eq{ue} and \eq{p2} to find
\beq
\eta ^{bc}(1-\C)_{\uc}{}^{\ub}(\C_{\una})_{\ua\ub} E_{c}{}^{\uc}
{\cE}_{b}{}^{\una}=0\ .
\la{Dirac3}
\eeq
We recall that ${\cE}_{b}{}^{\una} =  e_b{}^m {\cE}_m{}^{\una}$
and that
$E_{c}{}^{\uc}= m_c{}^d e_d{}^n {\cE}_n{}^{\uc}$.
Using these relations, the Dirac equation can be written as
\be
{\cE}_a(1-\C)\C^b m_b{}^a=0\ , \la{Dirac4}
\ee
where $\C^b=\C^m e_m{}^a$ and the target space spinor indices are suppressed.

The Dirac equation obtained above has a very similar form to those of
D-branes in ten dimensions \cite{d0,d1,d2,d3,d4}, and indeed we expect that
a double dimensional reduction would yield the 4-brane Dirac equation.

The emergence of the projection operator $(1-\C)$ in the Dirac equation
in the case of D-branes, and the other known super p-branes is due to the 
contribution of Wess-Zumino terms in the action (see, for example, ref. \cite{bst2}
for the eleven dimensional supermembrane equations of motion).
These terms are also needed for the $\kappa$-symmetry of the action. It
is gratifying to see that the effect of Wess-Zumino terms is
automatically included in our formalism through a geometrical route that
is based on considerations of the embedding of a world superspace into
target superspace.

\subsection{The Scalar Equation}

By scalar equation we mean the equation of motion for $x^{\unm}(x)$,
i.e. the coordinates of the target space, which are scalar fields from
the worldsurface point of view. In a physical gauge, these
describe the five scalar degrees of freedom that occur in the
worldsurface tensor supermultiplet.

The scalar equation is the leading component of the superspace equation
\eq{dil2} which we repeat here for the convenience of the reader:
\beq
\eta^{ab}\nabla_a E_b{}^{\una} E_{\una}{}^{b'} =
-\ft18 (\C^{b'a})_{\c'}{}^{\b}\, Z_{a\b}{}^{\c'}. \la{sa}
\eeq
The superspace equation for the covariant derivative
\beq
\nabla_a=E_a{}^m \nabla_m +E_a{}^{\m}\nabla_{\m},
\eeq
when evaluated at $\th=0$ involves the worldsurface gravitino
$E_a{}^{\mu}(x)$ which is expressible in terms of the basic fields of
the worldsurface tensor multiplet. Since it is fermionic it follows that
the second term in the covariant derivative will be bilinear in fermions
(at least), and we shall henceforth drop all such terms from the
equations in order to simplify life a little. We shall temporarily make
a further simplification by assuming that the target space is flat. The
tensor $Z$, as we saw earlier, has two types of contribution, one
($T_a$) involving $H_{\una\unb\unc\und}$, and the other ($K_a$)
involving only terms which are bilinear or higher order in fermions. In
accordance with our philosophy we shall henceforth ignore these terms.

To this order the right-hand side of the scalar equation
vanishes as does the right-hand side of the tensor equation \eq{tensor}.
Multiplying the scalar equation \eq{sa} with $E_{b'}{}^{\unc}$,
we see that it can be written in the form
\beq
\h^{ab}(\nabla_a E_b{}^{\unc}-  K_{ab}{}^c E_c{}^{\unc})=0\ ,
\eeq
where $K_{ab}{}^c$ is defined below. Using the relation
$E_b{}^{\unc}=m_b{}^d u _d{}^{\unc}$ and the definition of $X_{ab}{}^c$
in \eq{xx} we find that
\beq
K_{ab}{}^c := \nabla_a E_b{}^{\und}E_{\und}{}^{c} =
(\hat \nabla_a m_b{}^d)(m^{-1})_d{}^c + X_{ab}{}^c\ .
\eeq
Using the relation
\beq
\eta^{ab}\hat\nab_a m_b{}^{c}=0\ ,
\la{dil3}
\eeq
which we will prove later, we conclude that
$\eta^{ab} K_{ab}{}^c = \eta^{ab} X_{ab}{}^c$ .
As a result, we can express the scalar equation of motion in the form
\beq
\h^{ab}\hat\nab_a E_b{}^{\unc}=0\ .
\eeq
where $ \hat\nab_a E_b{}^{\unc}= \nab_a E_b{}^{\unc} - X_{ab}{}^d E_d{}^{\unc}$.
The relation \eq{dil3} allows us to rewrite the scalar equation of motion
in the form
\beq
m^{ab}\hat\nab_a u_b{}^{\unc}=0\ . \la{mdu}
\eeq

The next step is to find a explicit expression for the spin connection
$\hat\o_{a,b}{}^{c}$ associated with the hatted derivative. Using the
definition of $X_{ab}{}^d$ given in \eq{xx}, we find that this spin
connection is given by
\beq
\hat\o_{a,b}{}^{c}= \O_{a,b}{}^{c} +X_{a,b}{}^{c}=
E_a{}^m(\del_m u_b{}^{\unc})u_{\unc}{}^c\ .
\eeq
Recalling \eq{ue} and \eq{e}, we find that the hatted spin connection takes
the form
\beq
\hat\o_{a,b}{}^{c}=  m_a{}^f e_f{}^n\left(\partial_n e_b{}^m g_{mp}e^{cp}
+ e_b{}^m \partial_n {\cE}_m{}^{\und}\,{\cE}_{p\und}\,e^{cp}\right)\ .
\eeq
From this expression it is straightforward to derive the following
result; given any vector $V_m$ one has 
\beq
\hat\nab_a V_b=m_a{}^d e_d{}^n e_b{}^m \nabla_n V_m
\eeq
where
\beq
\nabla_n V_m= \partial_n V_m -\Gamma_{nm}{}^p V_p
\eeq
and
\beq
\Gamma_{nm}{}^p= \partial_n {\cE}_m{}^{\unc}{\cE}_{s\unc}\, g^{sp}\ .
\eeq
It is straightforward to verify that to the order to which we are working
this connection is indeed the Levi-Civita connection for the
induced metric $g_{mn}$.
\par
We are now in a position to express the scalar equation in its simplest form
which is in a coordinate basis using the hatted connection. Using the
above result we find that  \eq{mdu} can be written as
\beq
G^{mn}\nab_m{\cE}_n{}^{\una}=0.
\la{dil4}
\eeq

It remains to prove \eq{dil3}. Using the expression for $m_a{}^b$ given in
\eq{mmm}
we find that
\bea
\eta^{ab}\hat\nab_a m_b{}^{c} &=&-2 \hat\nab^b(h_{bde} h^{cde}) \nn\\
&=& -2h_{bde} \hat\nab^b h^{cde}=-\ft23 h_{bde}\hat\nab^c h_{bde}=-\ft13
\hat\nab^c
(h_{bde} h^{bde})=0\ .
\eea
In carrying out the above steps we have used the $h_{abc}$ equation of motion
and the self-duality of this field.

In the case of a non-flat target space the derivation is quite a bit longer
and the steps will be discussed elsewhere. One finds that the right hand side of
the scalar equation in the form of \eq{sa} is given by
\beq
\eta^{ab}\nabla_a E_b{}^{\una} E_{\una}{}^{c^\prime} =
-\ft1{144} (1-\ft23 \tr k^2) \epsilon^{ c^\prime e_1^\prime e_2^\prime
e_3^\prime e_4^\prime}
H_{e_1^\prime e_2^\prime e_3^\prime e_4^\prime}
+\ft23 m_a{}^b H^{c'}{}_{bcd} h^{acd}\ ,
\eeq
where
\beq
k_a{}^b := h_{acd}h^{bcd}\ .
\eeq
Using the steps given above this result can be expressed in the form
\bea
G^{mn}\nab_m{\cE}_n{}^{\unc} &=& {1\over \sqrt{-g}} (1-\ft23 \tr\,k^2)
\epsilon ^{m_1\cdots m_6 }\Big({1\over 6^2.4.5}
H^{\underline a}{}_{m_1\cdots m_6} \nn\\
&&+ \ft23 H^{\una}{}_{m_1m_2m_3}\,H_{m_4m_5m_6}\,\Big)
(\d_{\una}{}^{\unc}-{\cE}_{\una}{}^m{\cE}_m{}^{\unc})\ , \la{sfin}
\eea
where the target space indices on $H_4$ and $H_7$ have been converted
to worldvolume indices with factors of ${\cE}_m{}^{\una}$  and
\beq
H_{\underline d_1\ldots \underline d_4}= {1\over 7!}
\epsilon _{\underline d_1\ldots \underline d_4
\underline e_1\ldots \underline e_7}
H^{\underline e_1\ldots \underline e_7}\ .
\eeq
where $H_7$ is the seven form that occurs in the dual formulation of
eleven dimensional supergravity. One can verify that the ratio between
the two terms on the right hand side is precisely what one expects were
this term to have been derived from the expected gauge invariant
Wess-Zumino term of the form $C_6+4 C_3\wedge H_3$. We also
note that the last factor in \eq{sfin} implies that the RHS of the
equation vanishes identically when multipled with ${\cE}_{\unc}{}^q$, as
it should, indicating that only five of the eleven equations, which
correspond to the Goldstone scalars, are independent.

\subsection{ The Tensor Equation}

The tensor equation can be manipulated in a similar fashion.
If we consider the simplest case of ignoring the fermion bilinears and
assuming the target space to be flat we have, from \eq{tensor2}
\beq
\h^{ab} \hat\nab_a h_{bcd}=0\ .
\eeq
We can relate $h$ to $H$ using \eq{hh} and take the factor of $m$ past the
covariant derivative using \eq{dil3} to get
\beq
m^{ab}\hat\nab_a (e_b{}^m e_c{}^n e_c{}^p H_{mnp})=0\ .
\eeq
Using similar steps to those given in the proved in the previous
subsection and converting to a coordinate basis we find the desired form
of the tensor equation in this approximation, namely
\beq
G^{mn}\nab_m H_{npq}=0\ .
\eeq
In the case of a non-trivial target space a lengthy calculation
is required to find the analogous result. One first finds that
\bea
\hat \nabla ^c h_{abc} &=& \ft1{288} m_a^f m_b^g
\epsilon _{fg e_1 e_2 e_3 e_4} H^{e_1 e_2 e_3 e_4}
- \ft1{72} \epsilon _{ab d e_1 e_2 e_3 } m^d_f H^{f e_1 e_2 e_3}\nn\\
&&+6 h^{e_1e_2}{}_{[a}h_{bc]}{}^{e_3}m^{cc_1}H_{c_1 e_1 e_2 e_3 }
+\ft43 h_{abc}h^{e_1 e_2 e_3}m^{c c_1}H_{c_1 e_1 e_2 e_3}\equiv Y_{ab}\ .
\eea
It is possible to rewrite $Y_{ab}$ in the form
\beq
Y_{ab}= {( \tilde K + m \tilde K + \ft14 mm \tilde K)}_{ab}
\eeq
where  $\tilde K _{ab}= -{1\over 36.4!} \epsilon _{abcdef}H^{cdef}$,
$(m\tilde K)_{ab}= m_{[a}^c \tilde K_{b]c}$,
$(mm\tilde K)_{ab}= m_a^c m_b^d \tilde K_{cd}$.
The scalar equation of motion can also be  expressed in the form
\beq
G^{mn}\nab_m H_{npq}= {1\over (1-\ft23 \tr\,k^2)}
e_p^a e_q^b (4Y+4mY +mmY)_{ab}\ ,
\eeq
where $m Y $ and $mm Y $ are defined in a similar way
to the $m\tilde K$ and $mm\tilde K$  terms above.

\subsection{The $\kappa$-Symmetry Transformations}

The $\kappa$-symmetry transformations are related to odd worldsurface
diffeomorphisms. Under an infinitesimal worldsurface diffeomorphism
$\d z^M=-v^M$ the variation of the embedding expressed in a preferred frame
basis is
\beq
\d z^{\unA}\equiv \d z^{\unM} E_{\unM}{}^{\unA}= v^A E_A{}^{\unA}\ .
\eeq
For an odd transformation ($v^a=0$) one has
\bea
\d z^{\una} &=&0\ , \nn\\
\qquad \d z^{\ua} &=& v^{\a} E_{\a}{}^{\ua}\ . \la{ks1}
\eea
The vanishing of the even variation $\d z^{\una}$ is typical of
$\k$-symmetry and follows from the basic embedding condition \eq{mc1}.

The relation between the parameter $v^\a$ and the familiar $\k$
transformation parameter  $\k^{\ua}$ can be expressed as
\beq
v^\a = \k^{\uc} E_{\uc}{}^\a\ .
\eeq
Therefore, recalling \eq{p1}, the $\k$ transformation rule \eq{ks1} takes the
form
\beq
\d z^{\ua}= \k^{\uc} (1+\C)_{\uc}{}^{\ua}\ , \la{ks2}
\eeq
where we have absorbed a factor of two into the definition of $\k$. It
is understood that these formulae are to be evaluated at $\theta=0$, so that
they are component results.

There remains the determination of the $\k$-symmetry transformation of
the antisymmetric tensor field $B_{mn}$. It is more convenient to
compute the $\k$ transformations rule for the field $h_{abc}(x)$. (The
relation between the two fields is described earlier.) Thus we need to consider
\beq
\d h_{abc}=\k^{\uc} E_{\uc}{}^\a \nabla_\a h_{abc}\
\qquad {\rm at}\ \th=0\ . \la{dh1}
\eeq
By including a Lorentz transformation we may write this transformation as
\beq
\d h_{abc}= \k^{\uc} E_{\uc}{}^\a \hat \nabla_\a h_{abc}\ .
\eeq
We have calculated $\hat \nabla_\a h_{abc}$, and the derivation of the 
result will be given elsewhere \cite{hsw}. Using this result, we find
\be
\d h_{abc}= -\ft{i}{16} m_{[a|}{}^d\,{\cE}_d(1-\C)\C_{|bc]} \k\ , \la{dhf}
\ee
where $\C_a=\C^m e_{ma}$ and the target space spinor indices are suppressed. One can 
check that the RHS is self-dual, modulo the Dirac equation \eq{Dirac4}.

\section{Double Dimensional Reduction }

The procedure we shall adopt now is to use double-dimensional reduction \cite{ddr}
to obtain a set of equations for a 4-brane in ten dimensions and then to
compare this set of equations with the equations that one derives by
varying the Born-Infeld action. We shall take the target space to be
flat and we shall ignore the terms bilinear in fermions on the
right-hand-side of \eq{dil} and \eq{tensor}, that is we drop the terms in
these equations that involve the quantity $Z$ defined by \eq{z} and we
also ignore terms involving the worldsurface gravitino. From the
previous section, we read off the resulting equations of motion:
\beqa
&& G^{mn} \nabla_m {\cE}_n{}^{\una} =0 \ ,\\
&&G^{mn} \nabla_m H_{npq}=0\ .
\eeqa

We can further simplify matters by considering the corresponding bosonic
problem, i.e. by neglecting $\underline{\th}$ as well. In this limit,
and recalling that we have assumed that the target space is flat, one
has
\beq
{\cE}_m{}^{\una} \quad \rightarrow \quad  \del_m x^{\una}.
\eeq

In order to carry out the dimensional reduction we shall, in this
section, distinguish 6 and 11 dimensional indices from 5 and 10
dimensional indices by putting hats on the former. We have
\beq
x^{\hat m}=(x^m,y)
\eeq
and
\beq
x^{\hat{\unm}}=(x^{\unm},y)
\eeq
so that the sixth dimension of the worldsurface is identified with the
eleventh dimension of the target space; moreover, this common dimension
is taken to be a circle, and the reduction is effected by evaluating the
equations of motion at $y=0$. The metric is diagonal:
\beq
g_{\hat m\hat n}=(g_{mn},1),
\eeq
and the sechsbein can be chosen diagonal as well:
\beq
e_{\hat m}{}^{\hat a}=(e_m{}^a, 1)\ , \la{ee}
\eeq
where both the five-dimensional metric and its associated f\"{u}nfbein
are independent of $y$. Since the fields do not depend on $y$, and since
the connection has non-vanishing components only if all of its indices
are five-dimensional, the equations of motion reduce to
\beqa
G^{mn}\nabla_m\del_n x^{\una} &=&0 \label{dx1}  \label{dx1}\\
G^{mn}\nabla_m F_{np} &=&0, \label{df1}
\eeqa
where
\beq
F_{mn}=H_{mny}\ . \la{fh1}
\eeq
Since $h$ in six-dimensions is self-dual, and since $H$ is related to
$h$ it follows that we only need to consider the $py$ component of the
tensor equation. It will be convenient to rewrite these equations in an
orthonormal basis with respect to the five-dimensional metric; this
basis is related to the coordinate basis by the f\"{u}nfbein. Using
$a,b$, etc. to denote orthonormal indices, the equations of motion
become
\beqa
G^{ab}\nabla_a\del_b x^{\una}&=&0 \\
G^{ab}\nabla_a F_{bc}&=&0,
\eeqa
where
\beq
G^{ab}=(\hat m^2)^{ab}\ , \la{gmm}
\eeq
and where we have introduced a hat for the six-dimensional $m$-matrix
for later convenience.

The claim is that these equations are equivalent to the equations of
motion arising from the five-dimensional Born-Infeld Lagrangian,
\beq
{\cL}=\sqrt{- \det K}
\eeq
where \beq K_{mn}=g_{mn} + F_{mn},
\eeq
$g_{mn}$ being the induced metric. To prove this we first show that the
Born-Infeld equations can be written in the form
\beqa
L^{mn}\nabla_m\del_n x^{\una}&=& 0 \la{dx2}\\
L^{mn}\nabla_m F_{np} &=& 0, \label{df2}
\eeqa
where
\beq
L=(1-F^2)^{-1}.
\eeq
When matrix notation is used, as in the last equation, it is understood
that the first index is down and the second up, and $F^2$ indicates that
the indices are in the right order for matrix multiplication. $L^{mn}$
is then obtained by raising the first index with the inverse metric as
usual. To complete the proof we shall then show that $G$ is proportional
to $L$ up to a scale factor.

The matrix $K$ is $1+F$ so that its inverse is
\beq
K^{-1}=(1+F)^{-1}=(1-F)L,
\eeq
from which we find
\beq
(K^{-1})^{(mn)}=L^{mn}\qquad (K^{-1})^{[mn]}=-(FL)^{mn},
\eeq
the right-hand side of the second equation being automatically
antisymmetric. Varying the Born-Infeld Lagrangian with respect to the
gauge field $A_m$, ($F=dA$), gives
\beq
\del_n(\sqrt{-\det K}(K^{-1})^{[mn]})=0.
\eeq
Carrying out the differentiation of the determinant, switching to
covariant derivatives, and using the Bianchi identity for $F$, one finds
\beq
\nabla_n (K^{-1})^{[mn]}+(K^{-1})^{[pq]}\nab_p F_{qn} (K^{-1})^{[mn]}=0.
\label{dk}
\eeq
Using the identity
\beq
(K^{-1})^{[mn]} F_{np}=\d_p{}^m-L_p{}^m
\eeq
and the expression for $(K^{-1})^{[mn]}$ in terms of $L$ and $F$ one
derives from \eq{dk}
\beq
L_n{}^q\nab_q(F^{np}L_p{}^m)+F^{pn}L_n{}^q\nab_q L_p{}^m=0.
\eeq
On differentiating the product in this equation one finds that the two
terms with derivatives of $L$ vanish by symmetry. Multiplying the
remaining term by $(L^{-1})_m{}^r$ then yields the claimed result,
namely \eq{df2}. A similar calculation is used to derive \eq{dx2}.

To complete the proof we need to show that $G^{mn}$ is proportional to
$L^{mn}$. We begin by setting
\beqa
f_{ab}&=&h_{ab5}\ , \la{fh2}\\
F_{ab}&=&e_a{}^me_b{}^n F_{mn}\ .
\eeqa
We then find
\beqa
h_{abc}&=&\ft1{2}\e_{abcde} f^{de}\ , \la{hf}\\
F_{ab}&=&(m^{-1})_a{}^c f_{cb}\ , \la{ff}
\eeqa
where $m_a^b=\hat m_a{}^b$. The first equation follows from the self-duality of
$h_{abc}$, while the second equation follows from \eq{hh}, \eq{ee},
\eq{fh1} and
\eq{fh2}.

We set
\beqa
\hat m_{\hat a}{}^{\hat b}&=&(\hat m_a{}^b,\hat m_a{}^5,\hat m_5{}^b,\hat
m_5{}^5)\\
&=&(m_a{}^b,M_a,M^b,N).
\eeqa
Recalling that
\beq
\hat m_{\hat a}{}^{\hat b}=(1-2h^2)_{\hat a}{}^{\hat b}
\eeq
one finds
\beqa
m_a{}^b &=& \d_a{}^b(1-2t_1)+ 8(f^2)_a{}^b \\
M_a     &=& -\e_{abcde} f^{bc} f^{de} \\
N       &=& (1+2t_1),
\eeqa
where $t_1={\rm tr}(f^2)$. Noting that $f_a{}^b M_b=0$,
as can be seen by symmetry arguments, it follows from \eq{ff} that
\beq
F_a{}^b M_b=0\ .\label{FM}
\eeq
Now, by a direct calculation, starting from \eq{gmm} one finds that
\beq
G_{ab}=A\h_{ab} + 16(f^2)_{ab}\ , \la{gaff}
\eeq
where
\beq
A=1-4t_1-4(t_1)^2 +16t_2\ ,
\eeq
and we have defined $t_2={\rm tr}(f^4)$. Now, multiplying
$G_{ab}=(m^2)_{ab}+M_aM_b$ with $(F^2)^{bc}$, and recalling \eq{ff}, one finds
\beq
GF^2=f^2\ .
\eeq
Using this relation in \eq{gaff} we find
\beq
G=A(1-16F^2)^{-1}\ .
\eeq

Therefore we have shown that (after a suitable rescaling of $F$), $G$ is
proportional to $L$ and hence the equations of motion arising from the
superspace formulation of the 5-brane, when reduced to a 4-brane in ten
dimensions, coincide with those that one derives from the Born-Infeld
Lagrangian.

\section{Conclusions}

The component form of the equations of motion for the 5-brane in
eleven-dimensions are derived from the superspace equations. They are
formulated in terms of the worldsurface fields $x^{\unn},
\theta^{\underline\m}, B_{mn}$. These equations are fully covariant in
six-dimensions; they possess six dimensional Lorentz invariance ,
reparametrization invariance, spacetime supersymmetry and $\kappa$
symmetry. We have also derived the $\kappa$ transformations of the
component fields. The fivebrane equations are derived from the
superspace embedding condition for p-branes which possess half the
supersymmetry found previously \cite{hs1} and used to find superspace
equations for the 5-brane in eleven-dimensions in \cite{hs2}. In the
superembedding approach advocated here, the $\kappa$-symmetry is nothing
but the odd diffeomorphisms of the worldsurface and as such invariance
of the equations of motion under $\kappa$-symmetry is guaranteed.

We have also carried out a double dimensional reduction to obtain the
4-brane in ten dimensions. We find agreement with the known
Born-Infeld formulation for this latter theory. The result in ten
dimensions which emerges from eleven dimensions appears in an unexpected
form and that generalises the Born-Infeld structure to incorporate the
worldsurface chiral 2-from gauge field.

In a recent paper \cite{f5} it was suggested that it was impossible to
find a covariant set of equations of motion for a self dual second rank
tensor in six dimensions. However, in this paper we have presented just
such a system whose internal consistency is ensured by the manner of its
derivation. We would note that although the field $h_{abc}$ which
emerges form the superspace formalism obeys a simple duality condition,
the field strength $H_{mnp}$ of the gauge field inherits a version of
this duality condition which is rather complicated. Using the solution
of the chirality constraint on the 2-form, we expect that our bosonic
equations of motion will reduce to those of \cite{f5,f6}.

In reference \cite{f7}, an auxiliary field has been introduced to write
down a 6D covariant action. It would be interesting to find if this
field is contained in the formalism considered in this paper. We note,
however, that in the approach of reference \cite{f7} one replaces the
nonmanifest Lorentz symmetry with another bosonic symmetry that is
equally nonmanifest, but necessary to eliminate the unwanted auxiliary
field and that the proof the new symmetry involves steps similar to
those needed to prove the nonmanifest Lorentz symmetry \cite{f7}.
Further, it is not clear if a 6D covariant gauge fixing procedure is
possible to gauge fix this extra symmetry.

In a forthcoming publication, we shall give in more detail the
component field equation and the double dimensional reduction
\cite{hsw}. We also hope to perform a generalized dimensional reduction
procedure to the worldsurface, but staying in eleven dimensions. In the
approach of this paper, there is little conceptual difference in whether
the worldsurface multiplet is a scalar multiplet (Type I branes), or
vector multiplets (D branes), or indeed tensor multiplets (M branes) and
we hope to report on the construction of all p-brane solutions from this
view point.

We conclude by mentioning some open problems that are natural to
consider, given the fact that we now know the 6D covariant field
equations of the M theory five-brane. It would be interesting to
consider solitonic p-brane solutions of these equations, perform a
semiclassical quantization, explore the spectral and duality properties
of our system and study the anomalies of the chiral system. Finally,
given the luxury of having manifest worldsurface and target space
supersymmetries at the same time, it would be instructive to consider a
variety of gauge choices, such as a static gauge, as was done recently
for super D-branes \cite{d2}, which would teach us novel and interesting
ways to realize supersymmetry nonlinearly. This may provide useful tools
in the search for the ``different corners of M theory''.

{\it Note Added}

While this paper was in the final stages of being written up, we saw two
related papers appear on the net \cite{f8, f9}. We hope to comment on the
relationship between these papers and the work presented in a subsequent
publication.

\pagebreak

\end{document}